\documentclass[prl, 10pt,twocolumn,superscriptaddress,floatfix,showpacs]{revtex4-2}

\usepackage[british]{babel}  
\usepackage[scaled=0.86]{berasans}  
\usepackage[colorlinks=true, allcolors=blue, urlcolor=blue]{hyperref}  
\usepackage{graphicx} 
\usepackage[babel]{microtype}  
\usepackage{amsmath,amssymb,amsthm,bm,amsfonts,mathrsfs,bbm} 

\usepackage{xspace}  
\usepackage{pgfplots}
\usepackage{xcolor,colortbl}
\usepackage{array}
\usepackage{bigstrut}
\newcommand{\ket}[1]{| #1 \rangle}
\newcommand{\bra}[1]{\langle #1|}
\newcommand{\ketbra}[1]{\ket{#1}\!\bra{#1}}
\newcommand{\ip}[2]{\langle #1|#2 \rangle}
\newcommand{\bea}{\begin{eqnarray}}
\newcommand{\eea}{\end{eqnarray}}
\newcommand{\tr}{\text{tr}}

\usepackage{dsfont}

\usepackage{xspace}  
\usepackage{pgfplots}
\usepackage{xcolor,colortbl}
\usepackage{array}

\newcommand{\x}{\mathrm{x}}
\newcommand{\y}{\mathrm{y}}
\newcommand{\I}{\mathrm{I}}

\begin{document}


\title{Sequential Quantum Maximum Confidence Discrimination }


\author{Hanwool Lee }
\affiliation{Faculty of Information Technology, University of Jyväskylä, 40100 Jyväskylä, Finland}
\affiliation{School of Electrical Engineering, Korea Advanced Institute of Science and Technology (KAIST), 291 Daehak-ro, Yuseong-gu, Daejeon 34141, Republic of Korea }

\author{ Kieran Flatt}
\affiliation{School of Electrical Engineering, Korea Advanced Institute of Science and Technology (KAIST), 291 Daehak-ro, Yuseong-gu, Daejeon 34141, Republic of Korea }

\author{Joonwoo Bae}
\affiliation{School of Electrical Engineering, Korea Advanced Institute of Science and Technology (KAIST), 291 Daehak-ro, Yuseong-gu, Daejeon 34141, Republic of Korea }

\begin{abstract}

Sequential quantum information processing may lie in the peaceful coexistence of no-go theorems on quantum operations, such as the no-cloning theorem, the monogamy of correlations, and the no-signalling principle. In this work, we investigate a sequential scenario of quantum state discrimination with maximum confidence, called maximum-confidence discrimination, which generalizes other strategies including minimum-error and unambiguous state discrimination. We show that sequential state discrimination with equally high confidence can be realized only when positive-operator-valued measure elements for a maximum-confidence measurement are linearly independent; otherwise, a party will have strictly less confidence in measurement outcomes than the previous one. We establish a tradeoff between the disturbance of states and information gain in sequential state discrimination, namely, that the less a party learn in state discrimination in terms of a guessing probability, the more parties can participate in the sequential scenario.

 \end{abstract}
 
 \maketitle

Fundamental principles of quantum information processing contain the no-go theorems that non-orthogonal quantum states cannot be perfectly copied \cite{Wootters:1982aa, RevModPhys.77.1225} nor discriminated \cite{Helstrom1969, Bergou2004, Barnett09, Bae2015}. Moreover, being closely connected, quantum correlations, such as entanglement and nonlocal correlations, are monogamous \cite{PhysRevA.61.052306, PhysRevLett.96.220503, PhysRevA.73.012112}, restricting quantum information processing across multiple parties. Peaceful coexistence of the no-go results can be observed in a sequential quatnum information task of multiple parties that apply non-destructive quantum operations, particularly weak measurements, by which the parties can sequentially extract nonlocal correlations \cite{PhysRevLett.114.250401}. How weak a measurement is determines the number of parties in the sequential scenario \cite{Colbeck2020}.

Sequential quantum state discrimination for an ensemble of two pure states has been shown such that many parties can sequentially perform unambiguous discrimination \cite{Bergou2013}, where conclusive outcomes do not give an incorrect guess. In other words, quantum channels are constructed between parties such that each party knows which states are to appear, thus can choose a measurement for unambiguous discrimination, and then passes a resulting state to the next party, which can also realize unambiguous discrimination. In this case, it is not weak measurements that make a quantum protocol sequential across parties; it is non-optimal discrimination where none of the parties attempt to minimize the probability of inconclusive outcomes. Apart from the case of two pure states, little is known so far. 

In this work, we establish the framework for sequential quantum state discrimination with a maximum-confidence (MC) measurement \cite{Croke2006}, which generalizes other strategies, such as unambiguous and minimum-error state discrimination. We present the construction of quantum channels between parties so that multiple parties can sequentially realize an MC measurement. Our findings show that all parties can achieve equally high confidence in the MC discrimination outcomes only when the positive operator-valued measure (POVM) elements corresponding to the conclusive outcomes of the MC measurement are linearly independent. Otherwise, sequential parties cannot maintain confidence in measurement outcomes; a party will have strictly less confidence than the previous ones. 

This work is organized as follows. We begin with sequential MC discrimination for two mixed states to clarify the sequential scenario and elucidate its structure. We establish the tradeoff between information gain in terms of guessing and state disturbance in the sequential scenario, from which we find that the strength of measurements determines the number of parties that can participate in the sequential scenario, similar to the tradeoff in sequential violations of Bell inequalities. We present the necessary and sufficient conditions for realizing sequential MC discrimination with equally high confidence. We also investigate sequential MC discrimination for trine qubit states that are linearly dependent. The strength of weak measurements can determine how many parties can participate in the sequential state discrimination. 

Before starting sequential state discrimination, we briefly review a quantum MC measurement \cite{Croke2006}. For an ensemble $\{q_{\x},\rho_{\x}\}_{\x=1}^n$, describing a state $\rho_{\x}$ given with a probability $q_{\x}$, an MC measurement provides the highest probability of making a correct guess about state preparation (P) once an outcome occurs in a measurement (M). Namely, it maximizes a conditional probability $\mathrm{Prob}_{P|M} (\x|\x)$, defined as confidence $C_{\x}$ on an outcome $\x$ via Baye's rule as follows,
\bea
C_{\x} := \max_{M } \frac{q_{\x} \mathrm{Prob}_{M|P} (\x|\x) }{\mathrm{Prob}_{ M} ( \x)} =  \max_{M_{\x}} \frac{ q_{\x} \tr[\rho_{\x} M_{\x} ]  }{ \tr[\rho M_{\x}] } \label{eq:conf1}
\eea
where $M_{\x}$ denotes a POVM element. An MC measurement with confidence $C_{\x}=1$ realizes unambiguous discrimination. It corresponds to minimum-error discrimination when confidence in all inconclusive outcomes is maximized on average and no inconclusive outcome occurs \cite{Barnett09}. One can find that a rank-one POVM element achieves maximization in Eq. (\ref{eq:conf1}); thus, all conclusive outcomes of an MC measurement are described by rank-one POVM elements \cite{Flatt2022, Lee2022}. 


Let us begin sequential discrimination with an ensemble of two quantum states with equal {\it a priori} probabilities. The instance with two states clarifies and elucidates the sequential scenario and the structure. We consider discrimination between two states,
\bea
\rho_{\x} & = & p \ketbra{\psi_{\x}}+\frac{1-p}{2} \I, ~\mathrm{where} \label{eq:st} \\
 \ket{\psi_{\x}} &= &\cos\frac{\theta}{2}\ket{0} - (-1)^{\x} \sin\frac{\theta}{2}\ket{1}.\nonumber 
 \eea
for $p\in (0,1]$ and $\x=1,2$. Note that unambiguous discrimination can apply only when $p=1$. To find an MC measurement for the ensemble, we exploit the semidefinite program and the optimality conditions \cite{Flatt2022, Lee2022} so that parameters satisfying the equalities in the following
\bea
C_{\x} \rho_{} = \frac{1}{2}\rho_{\x} + r_{\x} \sigma_{\x}~\mathrm{and}~ \tr[ M_{\x}   \sigma_{\x}] =0,~\mathrm{for}~\x=1,2 \label{eq:opt}
\eea
directly find an MC measurement, where $r_{\x}>0$ and $\sigma_{\x}$ is a state. The former is called Lagrangian stability, and the latter is called complementary slackness, for which we call $\sigma_{\x}$ a complementary state. It is straightforward to find $\{\sigma_{\x} = |\varphi_{\x}\rangle\langle \varphi_{\x}| \}$ where, for $\x=1,2$,
\bea
\ket{\varphi_{\x}} =   \left( \sqrt{ \frac{1+p\cos\theta }{{2}}}\ket{0} - (-1)^{\x }\sqrt{ \frac{1-p\cos\theta}{{2}}}\ket{1} \right). ~~~\label{eq:comp}
\eea
From the complementary slackness in Eq. (\ref{eq:opt}), we write by $ |{\varphi_{\x}^{\perp} } \rangle $ an orthogonal complement to the state above $\ket{\varphi_{\x}}$. Then, an MC measurement can be written as, 
\bea
M_{1}=c_{1} \ketbra{ {\varphi}_{2}^{\perp}}~\mathrm{and} ~M_{2}=c_{2} \ketbra{{\varphi}_{  1}^{\perp}}  \label{eq:optm2}
\eea
where $c_1,c_2 >0$, together with an additional POVM element $M_0 = \I - M_1 - M_2$ that describes the probability of inconclusive outcomes $p_0 = \tr[\rho M_0]$. A measurement outcome $\x$ concludes a state $\rho_{\x}$ with maximum confidence as follows, 
\bea
C_{\x} = \frac{1}{2}(1+\frac{p\sin\theta}{\sqrt{1-p^2\cos^2\theta}}) ~\mathrm{for}~\x=1,2. \label{eq:conf2}
\eea
For $p=1$, we have $C_1=C_2=1$, which corresponds to unambiguous discrimination. 

We now provide a channel that makes many parties sequentially perform MC measurements with equally high confidence in Eq. (\ref{eq:conf2}). The channel can be described by Kraus operators, $\mathcal{E}(\cdot) = \sum_i K_i (\cdot) K_{i}^{\dagger}$ where Kraus operators $ K_i = V_i \sqrt{M_i }$ are constructed from POVM elements in Eq. (\ref{eq:optm2}) for some unitary $V_i$, 
\bea \label{eq:mcmkraus}
K_1 & =& \sqrt{c_1} \ket{\phi_1}\bra{{\varphi}_{2}^{\perp} },~ K_2 = \sqrt{c_2} \ket{\phi_2}\bra{ {\varphi}_{1}^{\perp}}, ~\mathrm{and} \nonumber\\
K_0 &=& \sqrt{a_1}\ket{\phi_1}\bra{ {\varphi}_{2}^{\perp} }+\sqrt{a_2}\ket{\phi_2}\bra{ {\varphi}_{1}^{\perp} },\label{eq:kraus1}
\eea
where $|\phi_1\rangle$ and $|\phi_2\rangle$ are qubit states. Note that the construction is possible with parameters 
\bea
(a_{\x} + c_{\x})^{-1} = D^2(|{\varphi}_{1}^{\perp} \rangle, | {\varphi}_{2}^{\perp} \rangle ) \label{eq:condition2}
\eea
where $D(   | {\varphi}_{1}^{\perp} \rangle, | {\varphi}_{2}^{\perp} \rangle ) = \|   | {\varphi}_{1}^{\perp} \rangle \langle {\varphi}_{1}^{\perp} | -   | {\varphi}_{2}^{\perp} \rangle \langle { \varphi}_{2}^{\perp} |  \|_1 /2$ and the trace distance is defined as $\| A\|_1 = \tr\sqrt{A^{\dagger}A}$ for a self-adjoint operator $A$. With the channel in Eq. (\ref{eq:kraus1}), the next party can identify the resulting ensemble. One can find that a measurement,
\bea
M_{1}^{(2)}= c_{1} ^{(2)} \ketbra{ {\phi}_{2}^{\perp}}~\mathrm{and} ~M_{2}^{(2)} = c_{2}^{(2)} \ketbra{{\phi}_{  1}^{\perp} }  \label{eq:optm22}
\eea
gives outcomes with confidence equal to Eq. (\ref{eq:conf2}). Thus, we have shown that two parties can realize sequential MC discrimination with equally high confidence for two mixed states, to which unambiguous discrimination does not apply. 

In what follows, we write by $|\widetilde{\varphi}_{1}\rangle: = | \varphi_{2}^{\perp}\rangle$ and $|\widetilde{\varphi}_{2}\rangle: = | \varphi_{1}^{\perp}\rangle$ for convenience. The significance of a state $| \varphi_{2}^{\perp}\rangle$ lies in the fact that it can uniquely identify a state $|\varphi_1\rangle$ by rejecting the other one, $|\varphi_2\rangle$. This will apply to an ensemble of multiple states. For an ensemble $\{|\varphi_{\x} \rangle\}_{\x=1}^n$, a state $|\widetilde{\varphi}_{1}\rangle$ is defined such that $\langle \widetilde{\varphi}_{1} | \varphi_{\x}\rangle = 0$ for $\x\neq1$. Hence, a POVM element $|\widetilde{\varphi}_{1} \rangle\langle \widetilde{\varphi}_{1}|$ can uniquely identify a state $|{\varphi}_{1}\rangle$ by rejecting the others.

Then, a relation between sequential MC measurements exists; for measurements in Eqs. (\ref{eq:optm2}) and (\ref{eq:optm22}),
\bea
&&  \ip{ \widetilde{\phi}_1}{\widetilde{\phi}_2}  = T  \ip{ \widetilde{\varphi}_1}{ \widetilde{\varphi}_2}~\mathrm{where} \label{eq:T}\\
 &&T = \left[(1-c_1 D^2( | \widetilde{\varphi}_1 \rangle, | \widetilde{\varphi}_2 \rangle )) 
(1-c_2 D^2( | \widetilde{\varphi}_1 \rangle, | \widetilde{\varphi}_2 \rangle )) \right]^{-\frac{1}{2}}. \nonumber
\eea
The derivation is shown in Supplemental Material. Since $T>1$, it holds that $  \ip{ \widetilde{\phi}_1}{\widetilde{\phi}_2} >  \ip{ \widetilde{\varphi}_1}{ \widetilde{\varphi}_2}$, which may be taken into account in the construction of the channel in Eq. (\ref{eq:kraus1}). Note that the case with $p=1$ reproduces sequential unambiguous discrimination \cite{Bergou2013}. 

It is worth mentioning that the relation in Eq. (\ref{eq:T}) shows that while equally high confidence in measurements is preserved, quantum states are sequentially disturbed. In fact, considering all of the conclusive and inconclusive outcomes, one can find that a quantum channel does not increase the distinguishability of quantum states on average \cite{PhysRevLett.103.210401, RevModPhys.88.021002}, 
\bea
\| \rho_1 - \rho_2 \|_1 \geq \| \mathcal{E} (\rho_1 - \rho_2) \|_1, \label{eq:cont}
\eea
for states $\rho_1$ and $\rho_2$.

 \begin{figure}[t]
    \centering
    \includegraphics[scale=0.13]{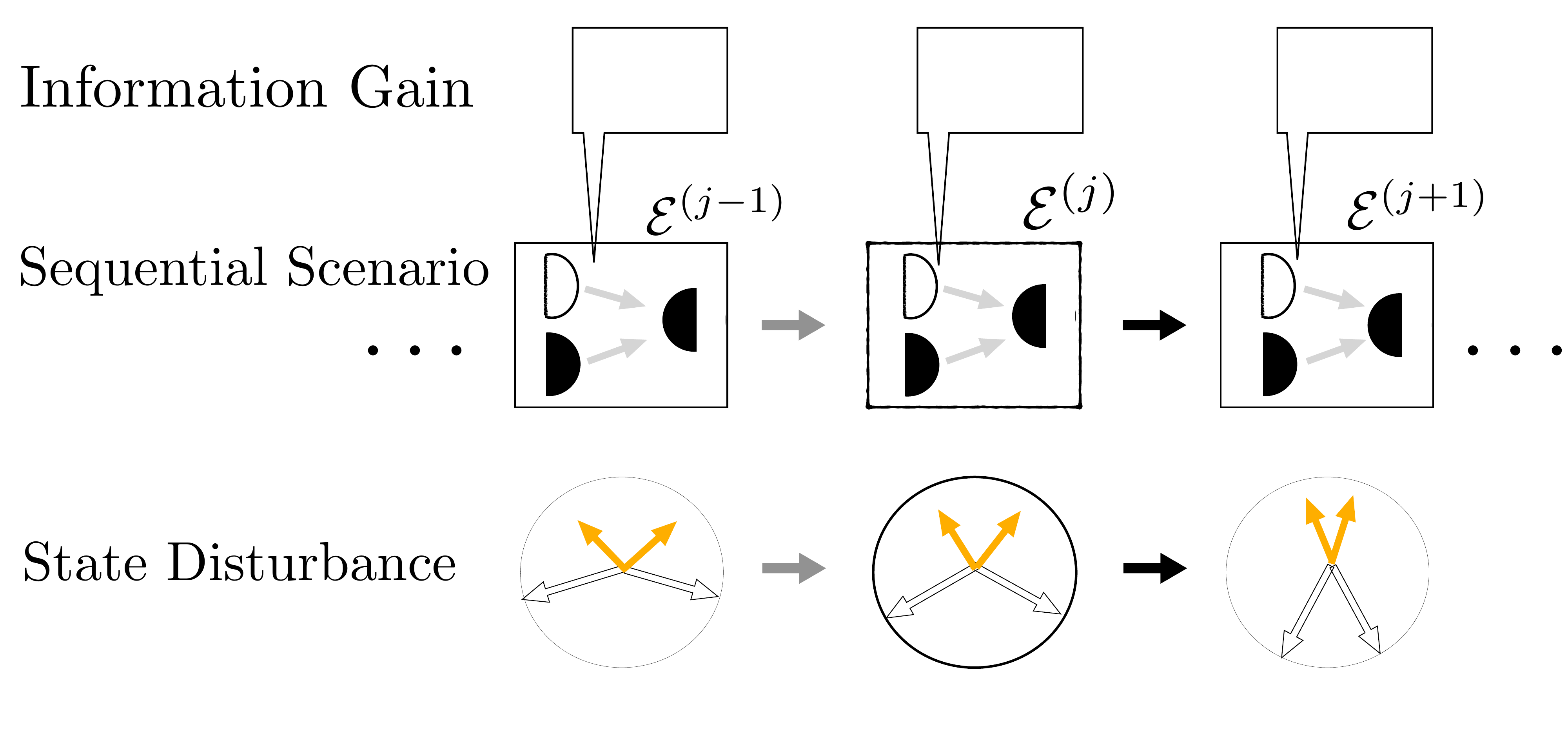}
    \caption{In the sequential scenario, the $j$-th party applies a quantum channel $\mathcal{E}^{(j)}$ that contains two types of outcomes, conclusive (white) and inconclusive (black) ones, see also Eq. (\ref{eq:kraus1}), and passes post-measurement states to the next party. Information gain is quantified by the guessing probability with conclusive outcomes in Eq. (\ref{eq:gue}). An MC measurement (white) disturbs states (orange), which evolve to be less distinguishable in the next party.   }
    \label{fig:enter-label}
\end{figure}

In the following, we show the tradeoff of information gain and state disturbance, see e.g., \cite{fuchs1996quantum}, in the sequential scenario of state discrimination. We quantify information gain by a guessing probability $G$ of a party, who learns about which state is given from conclusive outcomes,
\bea
G = C_1 \eta_1 + C_2 \eta_2.  \label{eq:gue}
\eea
where $\eta_{\x}$ is the probability of having an outcome $\x$ for $\x=1,2$.  For a guessing probability in Eq. (\ref{eq:gue}) an MC measurement can be devised such that it is least disturbing the distinguishability of states, quantified by the similarity between POVM elements, $s = | \langle \widetilde{\phi}_1  | \widetilde{\phi}_2 \rangle|$: 
\bea
\mathrm{minimize} &~& s = |\ip{ \widetilde{\phi}_1}{ \widetilde{\phi}_2}| \nonumber \\ 
\mathrm{subject~to} &~& G=\frac{1}{2}C(1-| \langle \widetilde{\varphi}_1 | \widetilde{\varphi}_2 \rangle |^2  )(c_1+c_2) \nonumber
\eea
where $G$ is given from Eq. (\ref{eq:gue}), where we have used $C=C_1=C_2$ from Eq. (\ref{eq:conf2}).  From the constraint in Eq. (\ref{eq:T}), it follows that $c_1=c_2$ and
\bea
s=  (1-\frac{G}{C} )^{-1}   | \langle \widetilde{\varphi}_1 | \widetilde{\varphi}_2 \rangle | \label{eq:s} 
\eea
for which the derivation of Eq. (\ref{eq:s}) is detailed in Supplemental Material. For instance, a channel for sequential MC discrimination in Eq. (\ref{eq:kraus1}) is least disturbing with states, 
\bea
\ket{{\phi}_{\x}}=\sqrt{\frac{1+s}{2}}\ket{0}- (-1)^{\x}\sqrt{\frac{1-s}{2}}\ket{1}\label{eq:postphi}
\eea
for $\x=1,2$, which can also be compared to the construction of sequential unambiguous discrimination \cite{Bergou2013}. From Eq. (\ref{eq:s}), it holds that
\bea
\frac{| \langle \widetilde{\varphi}_1 | \widetilde{\varphi }_2\rangle| }{ |\langle \widetilde{\phi}_1 | \widetilde{\phi}_2 \rangle |} = \frac{C-(C_1\eta_1 + C_2 \eta_2)}{C} = \eta_0 \label{eq:eta0}
\eea
where $\eta_0$ is the rate of inconclusive outcomes of the first party. Since $\eta_0<1$, POVM elements in sequential MC measurements become less distinguishable, meaning states in an ensemble are also less distinguishable on average, see Eq. (\ref{eq:cont}), since 
\bea
\sqrt{1- | \langle v_1 | v_2 \rangle|^2}=  \frac{1}{2}\| |v_1 \rangle \langle v_1 | -|v_2 \rangle \langle v_2| \|_1 \nonumber
\eea
for two states $|v_1\rangle$ and $|v_2\rangle$. The higher the rate of conclusive outcomes, the more disturbed states are.

Let us complete sequential MC discrimination for two states in Eq. (\ref{eq:st}) with the least disturbing measurements. We write POVM elements for an MC measurement of the $j$-th party 
\bea
M_{1}^{(j)}= c^{(j)}  |\widetilde{m}_{2}^{(j)}\rangle \langle \widetilde{m}_{2}^{(j)}|~\mathrm{and}~M_{2}^{(j)} =  c^{(j)} |\widetilde{m}_{1}^{(j)}\rangle \langle \widetilde{m}_{1}^{(j)} |  ~~~~ \label{eq:mcm}
\eea
and $M_{0}^{(j)} = \I  - M_{1}^{(j)} - M_{2}^{(j)}$ for inconclusitve outcomes. Each party is given an ensemble of two states given with {\it a priori} probabilities $1/2$,
\bea
S^{(j)} &= & \{\rho_{1}^{(j)}, \rho_{2}^{(j)} \}~~\mathrm{where} \label{eq:stj} \\
\rho_{1}^{(j)} & = & C   | {m}_{1}^{(j)}\rangle \langle  {m}_{1}^{(j)} | + (1-C )  | {m}_{2}^{(j)}\rangle \langle  {m}_{2}^{(j)} |~\mathrm{and} \nonumber \\
\rho_{2}^{(j)} & = & C   | {m}_{2}^{(j)}\rangle \langle  {m}_{2}^{(j)} | + (1-C )  | {m}_{1}^{(j)}\rangle \langle  {m}_{1}^{(j)} |\nonumber
\eea
where $\langle \widetilde{m}_{1}^{(j)} | {m}_{2}^{(j)}\rangle =0$ and $\langle \widetilde{m}_{2}^{(j)} | {m}_{1}^{(j)}\rangle =0$. It is clear that the party can have confidence $C$ on each conclusive outcome with a measurement in Eq. (\ref{eq:mcm}).

To make sequential discrimination possible, the state manipulation of the $j$-th party is characterized by Kraus operators as follows,
\bea
&& K_{1}^{(j)} = \sqrt{c^{(j)}}  \ket{ m_{1}^{(j+1)}}\bra{\widetilde{m}_{2}^{(j)} },~ K_{2}^{(j)} = \sqrt{c^{(j)}} \ket{m_{2}^{(j+1)}}\bra{ \widetilde{m}_{1}^{(j)}}, \nonumber\\
&& \mathrm{and}~  K_0 = \sqrt{ a_{1}^{(j)} }\ket{m_{1}^{(j+ 1)}}\bra{ \widetilde{m}_{2}^{(j)} } + \sqrt{a_{2}^{(j)}} \ket{ m_{2}^{(j+1)} }\bra{ \widetilde{m }_{1}^{(j)} }, ~~~~~~\label{eq:krauss}
\eea
where $a_{1}^{(j)} =a_{2}^{(j)}$ from Eq. (\ref{eq:condition2}), see also Supplemental Material for the detailed derivation. 

In addition, decompositions of states in Eq. (\ref{eq:st}) as follows,
\bea
\rho_1 &= & C| \varphi_1\rangle\langle \varphi_1 | + (1-C) | \varphi_2\rangle\langle \varphi_2 |~  \mathrm{and} \nonumber \\
\rho_2 & = & C| \varphi_2\rangle\langle \varphi_2 | + (1-C) | \varphi_1 \rangle\langle \varphi_1 |  \label{eq:decom}
\eea
elucidate the structure of sequential MC measurements. Then, all parties participating in sequential MC discrimination share the same value of $C$ in decompositions of their states, i.e., $\{ |m_{1}^{(j)}\rangle, |m_{2}^{(j)}\rangle\}$ for the $j$-th party. 

The tradeoff between information gain and state disturbance in Eq. (\ref{eq:eta0}) is generalized for each party in sequential MC measurements,
\bea
\eta_{0}^{(j)} = \frac{\widetilde{s}^{(j)} }{\widetilde{s}^{(j+1)}}~\mathrm{where}~\widetilde{s}^{(j)} = | \langle \widetilde{m}_{1}^{(j)} | \widetilde{m}_{2}^{(j)} \rangle  |
\label{eq:to}
\eea
The inner product of two POVM elements keeps increasing as more parties participate in the sequential scenario. With $R+1$ parties, it holds that 
\bea
\eta_{0}^{(1)} \times \eta_{0}^{(2)} \times \cdots \times \eta_{0}^{(R)} = \frac{\widetilde{s}^{(1)}}{ \widetilde{s}^{(R+1)}} \nonumber
\eea
We ask $\widetilde{s}^{(R+1)} < 1 -\delta$ for some $\delta >0$ so that the last party can also perform an MC measurement with equally high confidence. Assuming that rates of inconclusive outcomes are identical, $\eta_0 = \eta_{0}^{(j)}$ for all $j=1,\cdots, R$, we can limit the number of parties that can participate in the sequential scenario, 
\bea
R<  \left(\log  \frac{ \widetilde{s}^{(1)}}{1-\delta} \right) \frac{1}{ (\log\eta_0)} \label{eq:trR}
\eea
in terms of an initial condition $\widetilde{s}^{(1)}$ and inconclusive rate $\eta_{0}$. The higher the rate of inconclusive measurement outcomes, the larger the number of parties participating in sequential discrimination. Or, each party can control the rate of inconclusive outcomes so that more parties are allowed in sequential MC discrimination. 

We move to sequential MC measurements for more than two states. We first show that equally high confidence in measurements can be preserved only when POVM elements are linearly independent. Conversely, when POVM elements are linearly independent, we construct sequential measurements with equally high confidence. 

We consider an ensemble $S = \{q_{\x},\rho_{\x} \}_{\x=1}^n $ and its MC measurement, i.e., POVM elements $\{ M_{\x} \}_{\x=1}^n$, together with $M_0$ for inconclusive outcomes. It holds that for $\x=1,\cdots, n$ \cite{Flatt2022, Lee2022},
\bea
C_{\x}\rho - q_{\x} \rho_{\x} \geq 0~~\mathrm{and}~~\tr[ ( C_{\x}\rho - q_{\x} \rho_{\x}) M_{\x} ] =0. \label{eq:1}
\eea
Let $\{N_{\x} \}_{\x=0}^n$ denote an MC measurement for an ensemble resulting from a quantum operation $\mathcal{E}$ on the ensemble $S$. Assuming equally high confidence in conclusive outcomes, we have for $\x=1,\cdots,n$ 
\bea
\tr[ ( C_{\x}\rho - q_{\x} \rho_{\x}) \mathcal{E}^{\dagger} (N_{\x}) ] =0. \label{eq:2}
\eea
That is, POVM elements $\{ \mathcal{E}^{\dagger} (N_{\x})\}_{\x=1}^n$ form an MC measurement, meaning that $\mathcal{E}^{\dagger} (N_{\x}) = \alpha_{\x} M_{\x}$ for some $\alpha_{\x}>0$, which is possible if and only if POVM elements $\{ N_{\x}\}_{\x=1}^n$ are linearly independent \cite{Chefles1998}. 
 
In the other way around, suppose that the $j$-th party is with an ensemble of states its MC measurement as follows,
\bea
 S^{(j)} = \{q_{\x}^{(j)},\rho_{\x}^{(j)} \}_{\x=1}^n ~\mathrm{and} ~\{ M_{\x}^{(j)} \}_{\x=1}^n \nonumber
\eea 
as well as $M_{0}^{(j)}$ for inconclusive outcomes. Note that POVM elements for an MC measurement are rank-one \cite{Flatt2022, Lee2022}, which we write by 
\bea
M_{\x}^{(j)} = c_{\x}^{(j)} |\widetilde{m}_{\x}^{(j)} \rangle \langle  \widetilde{m}_{\x}^{(j)}| 
\eea
for $\x=1,\cdots, n$ and $M_{0}^{(j)} = \I - \sum_{\x=1}^n M_{\x}^{(j)}$. A channel from the $j$-th
 to the $(j+1)$-th parties is constructed with Kraus operators, with $ K_{\x}^{(j)} = V_{\x}^{j\rightarrow j+1} \sqrt{ M_{\x}^{(j)}}$ for some unitary $V_{\x}^{j\rightarrow j+1}$ that finds states $\{ |m_{\x}^{(j+1)} \rangle\}_{\x=1}^n$
\bea
 K_{\x}^{(j)} & = & \sqrt{c_{\x}^{(j)}} | m_{\x}^{(j+1 )}\rangle \langle \widetilde{m}_{\x}^{(j)} | ~\mathrm{for}~\x=1,\cdots,n \nonumber \\
\mathrm{and}~ K_{0}^{(j)} & = & \sum_{\x=1}^n \sqrt{a_{\x}^{(j)} } |m_{\x}^{(j+1 )} \rangle \langle \widetilde{m}_{\x}^{(j)}|. \label{eq:krj}
\eea
It is straightforward to find the ensemble of states of the $(j+1)$th party,
\bea
\rho^{(j+1)} & = & \sum_{\y}  \tr \left[ \rho_{}^{(j)} M_{\y}^{(j)} \right] |m_{\y}^{(j+1 )} \rangle \langle m_{\y}^{(j+1)} | +   \label{eq:jpe} \\
&& \sum_{k,l } \sqrt{a_k a_l} \tr \left[ \rho_{}^{(j)}   |\widetilde{m}_{ k}^{(j)} \rangle \langle \widetilde{m}_{l }^{(j)} | \right]   |m_{ l}^{(j+1)} \rangle \langle m_{k }^{(j+1)} |, \nonumber
\eea
and each state is given by
\bea
\rho_{\x}^{(j+1)} & = & \sum_{\y}  \tr \left[ \rho_{\x}^{(j)} M_{\y}^{(j)} \right]~  |m_{\y}^{(j+1)} \rangle \langle m_{\y}^{(j+1)} | +   \label{eq:jps} \\
&& \sum_{k,l } \sqrt{a_k a_l} \tr\left[ \rho_{\x}^{(j)}   |\widetilde{m}_{ k}^{(j)} \rangle \langle \widetilde{m}_{l }^{(j)} | \right]  |m_{ l}^{(j+1)} \rangle \langle m_{k }^{(j+1)} |.\nonumber
\eea
The $(j+1)$-th party can choose POVM elements for an MC discrimination, 
\bea
M_{\x}^{(j+1 )} = c_{\x}^{(j+1)} |\widetilde{m}_{\x}^{(j+1)} \rangle \langle  \widetilde{m}_{\x}^{(j+1)}| 
\eea
where $\langle \widetilde{m}_{\y}^{(j+1)} | {m}_{\x}^{(j+1)} \rangle =0$ for $\y\neq \x$. From Eqs. (\ref{eq:jpe}) and (\ref{eq:jps}), we compute,
\bea
C_{\x}^{(j+1)} & = & \frac{ q_{\x}\tr \left[ \rho_{\x}^{(j+1)}   M_{\x}^{(j+1 )} \right]}{\tr \left[\rho^{(j+1)}   M_{\x}^{(j+1 )} \right]} \nonumber \\
& = &  \frac{ q_{\x}  \tr \left[ \rho_{\x}^{(j)} M_{\x}^{(j)} \right] }{ \tr \left[ \rho_{}^{(j)} M_{\x}^{(j)} \right]  } =C_{\x}^{(j)}. \label{eq:eq}
\eea
Thus, both parties can have equally high confidence in a measurement. We summarize the result as follows.\\

{\it Proposition.} Sequential MC measurements with equally high confidence can be realized if and only if POVM elements are linearly independent. \\

Otherwise, when POVM elements of an MC measurement are linearly dependent, one may exploit a weak measurement to maintain confidence as high as possible in the sequential scenario. Let $\{ M_{\x}\}_{\x=1}^n$ denote POVM elements for conclusive outcomes of an MC measurement and $M_0$ for inconclusive ones. We construct a weak measurement,
\bea
N_{\x} & = & \epsilon M_{\x}~~\mathrm{for} ~\x=1,\cdots, n,~\mathrm{and}~ \nonumber \\
N_0 & = & (1-\epsilon)\I + \epsilon M_0 \label{eq:wem}
\eea
such that they realize an MC measurement. One can find a quantum channel accordingly,
\bea
K_{\x} =V_{\x} \sqrt{N_{\x}}~\mathrm{for}~\x=1,\cdots,n, ~\mathrm{and}~ K_{0} =V_{ 0 } \sqrt{N_{0}}, ~~~~~\label{eq:wkra} 
\eea
with some unitary $V_{\x}$. Let $C_{\x}^{(2)}$ denote confidence by an MC measurement $\{N_{\x}\}_{\x=1}^n$ and $C_{\x}^{(1)}$ by $\{M_{\x}\}_{\x=1}^n$, in which it is straightforward to find that
\bea
C_{\x}^{(2)}  = \left( 1- {\epsilon} \right)C_{\x}^{(1)} + O(\epsilon^2). \label{eq:general}
\eea
That is, we have $C_{\x}^{(2)} < C_{\x}^{(1)}$. The strength $\epsilon$ in a weak measurement can be chosen such that the next party can have an MC measurement with $\epsilon$-close to the confidence of the previous one, since $C_{\x}^{(2)}$ converges to $C_{\x}^{(1)}$ as $\epsilon\rightarrow 0$.
  
As an instance, we consider $n$ geometric uniform qubit states given with equal {\it a priori} probabilities $1/n$ \cite{PhysRevA.64.030303, 915636, PhysRevA.65.052308}, 
\bea
\{|\psi_{k}\rangle \}_{k=1}^n~~\mathrm{where}~ |\psi_k\rangle = \frac{1}{\sqrt{2}} \left(|0\rangle + e^{\frac{2\pi i }{n}k} |1\rangle\right) \nonumber
\eea
for which one can compute, $C_{\x} = 2/n$. It is also clear that POVM elements for an MC measurement 
\bea
M_{\x} = \frac{2}{n}|\psi_{\x}\rangle \langle \psi_{\x}| ~~\mathrm{for}~~\x=1,\cdots,n \nonumber
\eea 
are linearly dependent. From Eq. (\ref{eq:wkra}), a weak measurement by the $j$-th party constructs a quantum channel with Kraus operators parameterized by the rate of inconclusive outcomes $\eta^{(j)}_0 $
\bea
K^{(j)}_{\x} &=& \sqrt{\frac{2(1 - \eta^{(j)}_{0})}{n}} | \psi_{\x} \rangle \langle \psi_{\x} |~~\mathrm{and}~~ K^{(j)}_{0} = \eta^{(j)}_0 \I\nonumber
\eea
for $\x=1,\cdots,n$. We have for $j\geq 2$,
\bea
C_{\x}^{(j)} =  \frac{1}{2} \left( 1 + \prod_{k=1}^{(j-1)} \left(\frac{ 1 + \eta_0^{(k)}}{2} \right)   \right) C_{\x}^{(1)} \label{eq:sec} 
\eea
where $C_{\x}^{(1)} = 2/n$. One can find that the $(j+1)$-th party has strictly less confidence in measurement outcomes than the $j$-th one, i.e., $C_{\x}^{(j+1)} < C_{\x}^{(j)} $ for $j\geq 1$. The constraint on confidence such that $C_{\x}^{(j)}>C_{th}$ limits the number of parties $R$ that can participate in the sequential scenario. For instance, when $\eta_0:=\eta_{0}^{(k)}$ for all $k$, it holds,
\bea
R < 1 + \frac{\log (n C_{th} -1)}{\log \left((1+ \eta_0)/2\right)} \label{eq:RR}
\eea
which can be compared with Eq. (\ref{eq:trR}). The derivaion is detailed in Supplemental Material. Thus, parties in the sequential scenario can choose rates of inconclusive rate to weaken the measurement strength and attempt to maintain confidence as high as possible.

In conclusion, we have established a sequential scenario with MC measurements, a unifying framework of state discrimination including minimum-error and unambiguous state discrimination. We have shown the structure of sequential state discrimination by identifying POVM elements for MC measurements and characterized the tradeoff between information gain and state disturbance in the sequential scenario. We have presented the necessary and sufficient condition for sequential MC discrimination: sequential MC discrimination can be realized with equally high confidence in measurement outcomes if, and only if, POVM elements are linearly independent.  Our findings deepen the understanding of quantum information processing over trusted and cooperative parties. It would be interesting to apply our results to practical tasks such as sequential randomness \cite{Carceller2022} and quantum information tasks across multiple parties.


\section{acknowledgements}
H.L. is supported by Finnish Quantum Flagship from Research Council of Finland and BEQAH (Between Quantum Algorithms and Hardware) from Business Finland. K.F. and J.B. are supported by the National Research Foundation of Korea (Grant No. NRF-2021R1A2C2006309, NRF-2022M1A3C2069728, RS-2024-00408613, RS-2023-00257994) and the Institute for Information \& Communication Technology Promotion (IITP) (RS-2023-00229524).

\bibliography{bibliography.bib}
\bibliographystyle{apsrev4-1}

\newpage

\appendix

\section{SUPPLEMENTAL MATERIAL}

\section{Post-measurement states in sequential two-state discrimination: Derivation of Eq. (10) }

Recall that Kraus operators for a channel, $\mathcal{E}(\cdot) = \sum_i K_i (\cdot) K_{i}^{\dagger}$, are given by
\bea \label{eq:mcmkraus}
K_1 & =& \sqrt{c_1} \ket{\phi_1}\bra{{\varphi}_{2}^{\perp} },~ K_2 = \sqrt{c_2} \ket{\phi_2}\bra{ {\varphi}_{1}^{\perp}}, ~\mathrm{and} \nonumber\\
K_0 &=& \sqrt{a_1}\ket{\phi_1}\bra{ {\varphi}_{2}^{\perp} }+\sqrt{a_2}\ket{\phi_2}\bra{ {\varphi}_{1}^{\perp} }, \nonumber
\eea
with states $|\phi_1\rangle$ and $|\phi_2\rangle$. They form a valid POVM, 
\bea
\I&=&\sum_{\x=1,2} K_\x^\dag K_\x +K_0^\dag K_0 \nonumber \\
&=& (c_1+a_1)\ketbra{\varphi_2^\perp}+(c_2+a_2)\ketbra{\varphi_1^\perp} \nonumber \\
&~& +\sqrt{a_1 a_2}\ip{\phi_1}{\phi_2}(\ket{\varphi_2^\perp}\bra{\varphi_1^\perp}+\ket{\varphi_1^\perp}\bra{\varphi_2^\perp}).\nonumber 
\eea
Let us first take the inner product on both sides by the same state $\ket{\varphi_\x}$, for $\x=1,2$. We find that
\bea
c_\x+a_\x=\frac{1}{1-|\ip{\varphi_1}{\varphi_2}|^2}, ~\x=1,2. \nonumber 
\eea
By taking inner product on both sides by different states $\ket{\varphi_1}$ and $\ket{\varphi_2}$, and by using the above relation, we obtain
\bea
|\ip{\phi_1}{\phi_2}|=T|\ip{\varphi_1}{\varphi_2}| \nonumber 
\eea
where
\bea
T=[(1-c_1(1-|\ip{\varphi_1}{\varphi_2}|^2))(1-c_2(1-|\ip{\varphi_1}{\varphi_2}|^2))]^{-1/2} \nonumber
\eea

\section{Least Disturbing States: Derivation of Eq. (13)}
We approach the optimization by using the Lagrangian multiplier method. The given optimization is as follows:
\bea
\mathrm{minimize} &~& |\ip{ \widetilde{\phi}_1}{ \widetilde{\phi}_2}| =f(c_1,c_2)^{-\frac{1}{2}}|\ip{\widetilde{\varphi}_1}{\widetilde{\varphi}_2}|\nonumber \\ 
\mathrm{subject~to} &~& G=\frac{1}{2}C(1-| \langle \widetilde{\varphi}_1 | \widetilde{\varphi}_2 \rangle |^2  )(c_1+c_2) \nonumber
\eea
where we denote 
\bea
f(c_1,c_2)=(1-c_1 (1-| \langle \widetilde{\varphi}_1 | \widetilde{\varphi}_2 \rangle |^2))(1-c_2 (1-| \langle \widetilde{\varphi}_1 | \widetilde{\varphi}_2 \rangle |^2)) \nonumber
\eea
The Lagrangian is
\bea
\mathcal{L}& = &f(c_1,c_2)^{-\frac{1}{2}} |\ip{\widetilde{\varphi}_1}{\widetilde{\varphi}_2}|+ \nonumber\\
&& \lambda(G-\frac{1}{2}C(1-| \langle \widetilde{\varphi}_1 | \widetilde{\varphi}_2 \rangle |^2  )(c_1+c_2)) \nonumber
\eea
where $\lambda$ denotes a Lagrangian multiplier. By solving $\frac{\partial \mathcal{L}}{\partial c_\x}=0$ for $\x=1,2$,  we find that the optimal parameters satisfy $c_1=c_2$. 
Solving $\frac{\partial \mathcal{L}}{\partial \lambda}=0$, we find
\bea
c_1=c_2=\frac{G}{C(1-|\ip{\widetilde{\varphi}_1}{\widetilde{\varphi}_2}|^2)} \nonumber 
\eea
We finally find the optimal overlap,
\bea
|\ip{ \widetilde{\phi}_1}{ \widetilde{\phi}_2}| =(1-\frac{G}{C} )^{-1}   |\langle \widetilde{\varphi}_1 | \widetilde{\varphi}_2 \rangle|.  \nonumber
\eea

\section{Construction of Kraus Operators in a General Case: Derivation of Eqs. (18) and (25)}

For convenience, we consider $n+1$ Kraus operators so that $n$ POVM elements for conclusive outcomes are linearly independent, 
\bea
K_{\x}&=&\sqrt{c_x} \ket{\phi_x} \bra{ \widetilde{\varphi}_{\x}},~~\mathrm{for}~ \x=1,...,n \nonumber \\
\mathrm{and}~~K_{0}&=&\sum_{\x=1}^n \sqrt{a_{\x}} \ket{\phi_{\x}} \bra{ \widetilde{\varphi}_{\x}}. \nonumber
\eea
It holds that
\bea
\I&=&\sum_{\x=1}^n K_{\x}^\dag K_{\x} + K_0^\dag K_0 \nonumber \\
&=&\sum_{ \x=1}^n c_{\x} \ketbra{ \widetilde{\varphi}_{\x}}+\sum_{i,j=1}^n \sqrt{a_i a_j} \ip{\phi_i}{\phi_j} \ket{\widetilde{\varphi}_i} \bra{\widetilde{\varphi}_j},  \nonumber
\eea
from which,
\bea
\ip{ {\varphi}_i}{ {\varphi}_j}=\sqrt{a_i a_j} \ip{\phi_i}{\phi_j} \ip{  {\varphi}_i}{\widetilde{ \varphi}_i}\ip{ \widetilde{\varphi}_j}{ {\varphi}_j}. \nonumber
\eea
The above may be rewritten as 
\bea
\sqrt{a_i a_j} \ip{\phi_i}{\phi_j}  = \frac{\ip{ {\varphi}_i}{ {\varphi}_j}}{\ip{  {\varphi}_i}{\widetilde{ \varphi}_i}\ip{ \widetilde{\varphi}_j}{ {\varphi}_j}} \nonumber
\eea
where the left-hand-side is the Gram matrix of an ensemble of linearly independent states $\{a_i, |\phi_i\rangle \}_{i=1}^n$ and the right-hand-side is given from a construction of sequential MC discrimination. 

Note relations of an ensemble $\rho = \sum_{i} a_i |\phi_i\rangle \langle \phi_i|$ and its Gram matrix $G$ via $M = \sum_i \sqrt{a_i} |\phi_i \rangle \langle i| $, which may have a singular-value-decomposition, $M= UDV^{-1}$ with unitaries $U$ and $V$ and a diagonal matrix $D$. Then, we have 
\bea
\rho = MM^{\dagger} = UD^2 U^{\dagger}~~\mathrm{and}~~G=M^{\dagger}M = VD^2 V^{\dagger}. \nonumber
\eea
Hence, given a Gram matrix, one can construct an ensemble of linearly independent states.

\section{Sequential Discrimination of Symmetric States }

We consider an ensemble of geometrically uniform (GU) qubit states given with equal {\it a priori} probabilities, 
\bea
\ket{\psi_k} = \frac{1}{\sqrt{2}} \left( \ket{0} + e^{i2\pi k/n} \ket{1} \right)~\mathrm{for}~k=1,\cdots,n. \nonumber
\eea
We construct a quantum channel $\mathcal{E}$ from a weak measurement as follows,
\bea
K_{\x} = \sqrt{\frac{2\epsilon}{n}} |\psi_{\x}\rangle \langle \psi_{\x}|,~\mathrm{for}~\x=1,\cdots,n~\mathrm{and}~K_0 = \sqrt{1-\epsilon}~\I\nonumber 
\eea
so that $\sum_{\x=0}^n K_{\x}^{\dagger}K_{\x} = \I$, and note also that
\begin{equation} \label{sumeq0}
\sum_{j=1}^{n} e^{i 2 \pi j / n} = 0~\mathrm{and}~ \sum_{j=1}^{n} \ketbra{\psi_k}  = \frac{n}{2} \I  \nonumber
\end{equation}
It is clear that a state remains identical for inconclusive outcomes. For a state $|\psi_{\x}\rangle$, we have 
\bea
\mathcal{E}(|\psi_{\x}\rangle \langle \psi_{\x}|) = \epsilon \frac{2 }{n} \sum_{j=1}^n  |\langle \psi_{\x}|\psi_j \rangle|^2 |\psi_{j}\rangle \langle \psi_{j}| + (1-\epsilon) |\psi_{\x}\rangle \langle \psi_{\x}|.\nonumber
\eea
We compute the first term in the right-hand-side,
\bea
&& \epsilon \frac{2}{n}\sum_{j=1}^{n} | \langle \psi_k | \psi_j \rangle |^2 \ketbra{\psi_j}  \nonumber \\
&=& \frac{\epsilon}{2} \I   + \frac{\epsilon}{2} \sum_{j=1}^{n} \left(  e^{ i 2 \pi(k-j) / n} + e^{ - i 2 \pi(k-j) / n}  \right) \ketbra{\psi_k}  \nonumber\\
&=&\epsilon \left( \frac{1}{2} \I   +  \frac{1}{4} \left( e^{ i 2 \pi k / n} \ket{1}\bra{0} +  e^{ - i 2 \pi k / n} \ket{0} \bra{1} \right) \right)\nonumber\\
& = &  \epsilon\left( \frac{3}{4} \ketbra{\psi_{\x}} + \frac{1}{4} \ketbra{ {\psi}_{\x}^{\perp} } \right) \nonumber
\eea
Then, the post-measurement state is obtained as follows,
\bea
\mathcal{E} (|\psi_{\x} \rangle\langle \psi_{\x}|) & = &  \epsilon \left( \frac{3}{4} \ketbra{\psi_{\x}} + \frac{1}{4} \ketbra{ {\psi}_{\x}^{\perp} } \right) + (1-\epsilon) \ketbra{\psi_{\x}}  \nonumber \\
& = &  (1-\frac{\epsilon}{4}) \ketbra{\psi_{\x}} + \frac{\epsilon}{4} \ketbra{ {\psi}_{\x}^{\perp} }. \label{eq:1r} 
\eea
Let $\vec{a}$ denote the Bloch vector of the resulting state, $\mathcal{E} (|\psi_{\x} \rangle\langle \psi_{\x}|)$. From the relation,
\bea
\tr (\mathcal{E} (|\psi_{\x} \rangle\langle \psi_{\x}|))^2 = \frac{1}{2} (1+ |\vec{a}|^2) \nonumber
\eea
one can find that 
\bea
|\vec{a}| = 1-\frac{\epsilon}{2} \nonumber
\eea
which shows that the resulting states are also GU mixed states. 

For convenience, we replace $1-\epsilon$ with a parameter $\eta_0$, the rate of inconclusive outcome rate, and rewrite Eq. (\ref{eq:1r}) as follows,
\bea
\mathcal{E} (|\psi_{\x} \rangle\langle \psi_{\x}|) & = &   p_+ \ketbra{\psi_{\x}} + p_- \ketbra{ {\psi}_{\x}^{\perp} }. \label{eq:1rr} 
\eea
with 
\bea
p_{\pm} = \frac{1}{2} \left( 1 \pm \frac{1}{2}(1 + \eta_0) \right). \nonumber
\eea
From these, the Bloch vector is also found as,
\begin{equation}
\vec{a} = (p_+ - p_-) \left( 0, \sin(\frac{2 \pi \x}{n }),  \cos(\frac{2 \pi \x}{n}) \right) \nonumber
\end{equation}
and it is clear that,
\begin{equation}
| \vec{a} | = p_+ - p_- = \frac{1 + \eta_0}{2}. \nonumber
\end{equation}
When no measurement is performed, all outcomes are inconclusive, i.e., $\eta_0=1$, and a state is not disturbed.

In a similar manner, we compute the resulting state after sequential MC measurements by $R$ parties. It is straightforward to find that the resulting state of the $j$-th party is given by,
\bea
&& \mathcal{E}^{(j)} \circ \cdots \circ \mathcal{E}^{(1)}  ( | \psi_{\x} \rangle \langle \psi_{\x} |) \nonumber \\
&=& \frac{1}{2} \left( 1 + \prod_{k=1}^{(j)} \left( \frac{1 + \eta^{(k)}_0 }{2} \right) \right) \ketbra{\psi_{\x}} \nonumber \\
&& +  \frac{1}{2} \left( 1 - \prod_{k=1}^{(j)} \left( \frac{1 + \eta^{(k)}_0 }{2} \right)  \right) \ketbra{ {\psi}_{\x}^{\perp} } \nonumber
\eea
where $\eta_0 ^{(k)}$ is the inconclusive outcome rate of the $k$-th measuring party. 

In the same manner as above we can evaluate the length of the associated Bloch vector to be
\begin{equation}
|a^{(j)}| = \prod_{k=1}^{(j)} \left(\frac{  1 + \eta^{(k)}_0   }{2}\right). \nonumber 
\end{equation}
It is clear that the length of a Bloch vector monotonically decreases as more parties participate in the sequential scenario. It also follows that
\bea
C_{\x}^{(j)} =   \frac{1}{2} \left( 1 + \prod_{k=1}^{(j-1)} \left( \frac{1 + \eta_0^{(k)}}{2}    \right) \right)   C_{\x}^{(1)}  \nonumber
\eea
which keeps decreasing as $j$ increases. Note also that we have $C_{\x}^{(1)}=2/n $ for $n$ GU qubit states.  

In addition, we consider $R$ parties that have the same inconclusive outcome rate $\eta_0$. We have that,
\bea
C_{\x}^{(R)} =   \frac{1}{n} \left( 1 +   \left(\frac{ 1 + \eta_0^{} }{2} \right)^{R-1}  \right)   \nonumber
\eea
We ask the $R$-th party has confidence in measurement outcomes over a threshold $C_{th}$, i.e.,  $C_{\x}^{(R)} > C_{th}$ from which, we find a tradeoff relation between the rate of inconclusive outcomes $\eta_0$ and the number of parties $R$
\bea
R < 1 + \frac{\log (n C_{th} -1)}{\log \left((1+ \eta_0)/2\right)}. \nonumber
\eea
Thus, $R$ parties can choose $\eta_0$ such that all of them have cofidence larger than a threshold $C_{th}$.

\end{document}